\documentclass[%
preprint,
 amsmath,amssymb,
 aps,
showkeys,
]{revtex4}

\usepackage{graphicx}
\usepackage{dcolumn}
\usepackage{bm}
\usepackage{amsmath,amssymb, amsfonts}
\usepackage{enumitem, url}
\usepackage{tikz}
\numberwithin{equation}{section}

\begin{document}

\preprint{APS/123-QED}





\title{Gradient Sensing via Cell Communication}
\author{Dallas Foster}
\affiliation{Department of Mathematics, Oregon State University, Corvallis, OR 97331, USA} 
\author{Brian Frost-LaPlante}
\affiliation{Department of Electrical Engineering, Columbia University, New York, NY 10027, USA}
\author{Collin Victor}
\affiliation{Department of Mathematics, University of Nebraska at Lincoln, Lincoln, NE 68588, USA} 
\author{Juan M. Restrepo}
\affiliation{Department of Mathematics, Oregon State University, Corvallis, OR 97331, USA}
\affiliation{Kavli Institute for Theoretical Physics, University of California, Santa Barbara, Santa Barbara, CA 93106, USA}

\date{\today}

\begin{abstract}
Experimental evidence lends support to the conjecture that the ability of chains of cells to sense the gradient of an external chemical concentration could rely on cell-to-cell communication. This is the basis for the gradient sensing nature of a specific model type of the Local Excitation, Global Inhibition (LEGI) principle, wherein the strength of the external chemical field is sensed through a comparison between a local exciting species and a global inhibitor that is shared via intra-cellular reactions in the cell chain\cite{Mugler2016}. 
 
In this study we generalize the nearest neighbor communication mechanism in the above-mentioned LEGI model in order to explore how the chemical sensing characteristics depend on the parameterization of the communication itself, cell size, and the radius of influence of neighboring cells. 

It was found that the radius of influence was less important than the approximating model for communication. Higher order approximations to the communication mechanism were better able to
sense an external gradient. However, an analysis of the signal to noise ratio established that higher order models for communication were more prone to noise and thus have a lower signal to noise ratio. The generalization as well as the tools used in the analysis of the dynamics can be extended to more heterogeneous networks and can thus prove useful in using models and observations in the process of understanding chemical gradient via LEGI models with a communication component.
\end{abstract}

\maketitle

\section{Introduction}
\label{sec:intro}
The motion of organisms and the biased growth in organisms as a consequence of exposure to a chemical gradient are two manifestations of what is called chemotaxis. Chemotaxis plays a role in wound healing, neuronal network development, and tumor metastasis. Determining how cells detect a chemical gradient, and the extent of this capability, are critical to understanding the individual and collective dynamics of cellular chemotaxis. 

Studies aimed at understanding chemotactic gradient sensing have traditionally focused on the physical capabilities of a single cell \cite{Berg1977, Parent1999, Levchenko2002}. It is now clear that limiting ourselves to understanding single-cell gradient sensing does not generally explain chemotaxis for collections of cells \cite{Camley2016, Ellison2016}, particularly when the cells in question do not have any apparent dedicated organelle that can sense spatial chemical biases \cite{Keller1970, Levchenko2002, Friedl2009, Theveneau2010}. 

The Local Excitation Global Inhibition (LEGI) framework proposes that the dynamics of a receptor-chemical reaction internal to a cell can model how cells sense the direction of an external chemical gradient \cite{Parent1999, Levchenko2002}. The LEGI framework has been extended to incorporate chemical interactions between chains of cells \cite{Malet-Engra2015, Friedl2009, Mugler2016}.
 
Cell-to-cell communication is a signaling pathway, however, the intercellular signal must compete with the ever-present noise. Experiments have demonstrated that it is the norm, rather than the exception, for the ambient chemical environment itself to be noisy and, as a result, interfere with the cell's ability to detect the chemical signal \cite{Endres2009, Hu2010, Jilkine2011a, Hu2011}. Berg and Purcell \cite{Berg1977} first made this point in analyzing the perception of a gradient by a single cell. However, a consequence of cell-to-cell communication is that internal, as well as intercellular, noise play a role \cite{Endres2009} and thus gradient sensing by chains of cells introduces complexities in the competition of noise and signal communication. Obviously, the details also depend critically on how communication is modeled.

Mugler \textit{et al.}\cite{Mugler2016}, using an adaptation of the LEGI approach \cite{Levchenko2002} with a simple nearest neighbor communication scheme, examined the gradient sensing capabilities of an interacting chain of cells. Their model includes a local reaction, coupled with an external diffusive chemical species and cell-to-cell communication. The background noise is parametrized by a stochastic process. They estimated the physical range over which cell chains are able to resolve a gradient by looking at the relative strength of the effect of communication to the noise in the system (see also \cite{Endres2009}).

The estimates in Mugler {\it et al.} are dependent on the specific choice of model for the communication mechanism. This paper generalizes the model advanced by Mugler {\it et al.}\cite{Mugler2016}, by introducing a broad framework for the parameterization of cell interaction. Doing so allows for a more systematic analysis of the impact of communication on the ability of cells to collectively detect gradients in a noisy chemical environment. The goals of this study are to specifically explore how the various models for communication affect the ability of cell collections to sense a gradient and thereafter propagate this signal through the network. In addition to the critical signal-to-noise (SNR) analysis for the general model, and several parameterizations including the nearest neighbor model of Mugler {\it et al.}, we examine fundamental physical characteristics of the global excitations like the phase/group velocities and space/time correlations.

\section{The Nearest-Neighbor LEGI Model (NNLEGI)}
\label{sec:LEGIMODEL}
The following evolutionary model due to Mugler {\it et al.} \cite{Mugler2016} captures the reaction equations of a one-dimensional chain of cells, exposed to a diffusing external chemical concentration. Its distinguishing feature is the presence of nearest-neighbor, cell-to-cell, chemical concentration communication, which is why we call it the Nearest-Neighbor LEGI (NNLEGI) model.

Each cell sports active receptors, $r$, whose chemical species is activated by the external concentration $c$ with rate $\alpha$ and activates the local and communicable species $x$ and $y$ at a rate $\beta$. $x$ is local to each cell whilst $y$ can be passed between cells, at a constant rate $\gamma$. Hence $x$ and $y$ are activated and deactivated at the same rate, but the exchange of $y$ between cells leads to differences in amounts of $x$ and $y$ in an individual cell. This difference between the local and global concentrations is proposed to be the underlying signal the cell uses to infer the chemical gradient.

We consider a chain of $m$ cells, each of length $\tilde{a}$. The total length of the cell chain is given by $L = m\tilde{a}$. 
We denote by $c_0$ the average value of the dimensional concentration $\tilde{c}$. Non-dimensionalized, the NNLEGI model is
  \begin{equation}\label{eq:DiscreteOriginalLEGI}
    \begin{split}
    \frac{\partial c}{\partial t} &= \nabla^2 c - \sum_{n=1}^{m}\delta\left(\frac{z}{L} - \frac{z_n}{L}\right)\frac{dr_n}{dt}, \\
    dr_n &= \alpha c_n dt - \mu r_n dt+ d\eta_n, \\
    dx_n &= \beta r_n dt - \nu x_n dt+ d \xi_n,\\
    dy_n &= \beta r_n dt- \nu \sum_{n'=1}^m M_{nn'}y_{n'} dt+ d \chi_n.
    \end{split}
  \end{equation} 
$M$ is a symmetric, nearest-neighbor coupling term that gives NNLEGI its name,
\begin{equation*}
     M_{nn'} = \delta_{n,n'}(1 + 2\gamma/\nu) - (\delta_{n-1,n'} + \delta_{n+1,n'})(\gamma/\nu).
\end{equation*}
The non-dimensionalized position is $z = \tilde{z}/L$, and time is $t = \tilde{t} D/L^2$ where $D$ is the diffusion coefficient of $c$. The $n^{th}$ cell is located at $z_n$. The non-dimensionalized rates $\alpha$, $\beta$, $\mu$, $\nu$, $\gamma$ are obtained by multiplying their dimensional counterparts by $\tilde{a}^2/D$. Similarly, each chemical concentration is inversely scaled by $c_0$. $ a = \tilde{a}/L = 1/m$ is the non-dimensional cell size. The concentration $c$ is modeled by a diffusive equation and, at each cell location, the external concentration is coupled to the internal concentration by a system or ordinary differential equations for $r, x$, and $y$.
Thermal fluctuations are captured by $d \eta_n, d \xi_n$, and $d \chi_n$, representing uncorrelated incremental noise processes with known variances. As argued in \cite{Mugler2016}, $d\eta_n = \alpha \overline{c}_n \delta F_n$, comes from the ``equilibrium binding and unbinding of external molecules to receptors" and is in terms of $F_n$, the free energy difference associated with a molecule unbinding from the $n$th cell. The incremental noise processes representing thermal effects obey
\begin{equation}
    \label{eq:noiseterms}
    \begin{split}
    &\scriptstyle\langle d\xi_n(t) d\xi_n'(t') \rangle = \delta_{nn'}\left(\beta \overline{r}_n + \nu \overline{x}_n \right)\delta(t-t'),\\
    &\scriptstyle\langle d\chi_n(t) d\chi_n'(t') \rangle = [ \delta_{nn'}\left(\beta \overline{r}_n + \nu \overline{y}_n + 2\gamma \overline{y}_{n} + \gamma \overline{y}_{n-1} + \gamma \overline{y}_{n+1} \right)\\
    &\scriptstyle - \delta_{n-1, n'}\left(\gamma \overline{y}_{n-1} + \gamma \overline{y}_{n+1}\right) - \delta_{n+1, n'}\left(\gamma \overline{y}_{n-1} + \gamma \overline{y}_{n+1}\right) ] \delta(t-t'),
    \end{split}
\end{equation}
where positive terms account for the Poisson noise and negative terms account for the anti-correlation from each exchange. $\overline{c}, \overline{r}, \overline{x}$, and $\overline{y}$ represent mean steady state solutions, which will be explicitly described in Section \ref{sec:SNR}.

\section{Generalization of the NNLEGI Model}
\label{sec:GeneralizationLEGI}

We consider the limiting generalization of (\ref{eq:DiscreteOriginalLEGI}) in the sense that the number of cells $m \rightarrow \infty $ and $ a \rightarrow 0$. Heuristically, this represents the limit of indistinguishable cells as the cell size is negligible compared to the size of the domain overall. In this limit, the resulting set of partial differential equations can be written as
    \begin{equation}
    \label{eq:ContinuousLEGI}
        \begin{split}
            \frac{\partial c}{\partial t} &= \nabla c - \int_\mathcal{C} \delta(z - z^*) \frac{dr}{dt} dz^* , \\ 
            dr &= \alpha c - \mu r + d\eta, \\
            dx &= \beta r - \nu x + d\xi, \\  
            dy &= \beta r - \nu y + \gamma (w \ast y) + d\chi,
        \end{split}
    \end{equation}
for $t>0$ and $z\in \Omega$, the physical domain. $r(z, t), x(z, t)$, and $y(z, t)$ represent the concentrations and are assumed known at $t=0$. 

The terms $d\eta(z,t), d\xi(z,t), d\chi(z,t)$ represent the spatially-continuous analogues of their counterparts in (\ref{eq:DiscreteOriginalLEGI}). We will refer to the last term in the $y$ equation of (\ref{eq:ContinuousLEGI}) as the {\it communication term}. It is the convolution of $y$ and a weight function $w(z)$. The (spatial) convolution is defined as 
    \begin{equation}\label{eq:ConvolutionDefinition}
        (w*y)(z,t) = \int_{-\infty}^{\infty} w(u)y(z-u,t)\;du.
    \end{equation}
    Symmetries, continuity and boundedness constraints need to be imposed on the kernel $w$ for this general communication term to make sense; other specifications of the model for the kernel $w$ and $\gamma$ would need to be obtained from the cells and the cell network being modeled. In what follows we will show that the simplest approximation of the communication term in fact leads to the NNLEGI model, however, the manner in which this generalized communication term is approximated leads to several other alternative LEGI models with communication.

\subsection{Alternative Communication Models}
\label{sec:ConvolutionGeneralization}

In order to investigate the impact of modeling chemical communication 
in (\ref{eq:ContinuousLEGI}) we focus on the convolution kernel $w$. 
When the diffusion scale is small compared to the correlation length of communication, the manner in which the kernel is mapped onto a discrete cell chain becomes important. Specifically, some approximations lead to sensitive dependence of gradient sensing estimates on the nature of the kernel.

Among the properties of relevance to the physics, translational symmetry dictates that the kernel $w$ has the property 
    \begin{equation*}
        \int_{\mathcal{C}} w(z) dz = 0.
    \end{equation*} 
Furthermore, $w$ should not endow directional preferences to the fluxes of concentration. Hence, $w$ should be spatially symmetric and therefore odd functions should lie in the kernel of $w$. 

We analyze the operator when the support of $w$ is small. Following \cite{Murray2002}, we expand $y$ into a Taylor series centered at $z$, obtaining, to lowest order, a second-order operator approximation. To see this,
    \begin{equation}\label{eq:TaylorExpansion}
        \begin{split}
            &\int_{-\infty}^{\infty} w(u)y(z-u,t)\;du \approx \\
            &\int_{-\infty}^{\infty} w(u)\Big(y(z,t) - u y'(z,t) + \frac{u^2}{2} y''(z,t)\Big)\;du.
        \end{split}
    \end{equation}
Symmetry and definiteness of $w$ implies that the first two terms in the preceding integral expression are zero. The remaining term is a number solely dependent on the structure of the kernel in use. We call this number $\Gamma$, allowing us to simply write
    \begin{equation*}
        w*y \approx y''(z,t)\int_{-\infty}^{\infty} \frac{u^2}{2}w(u)\;du = \Gamma y''(z,t).
    \end{equation*} 
Replacing this approximation into (\ref{eq:ContinuousLEGI}) we obtain the second order approximation
    \begin{equation} \label{eq:y2DApproximation}
        \frac{\partial y}{\partial t} = \beta r(z,t) - \nu y(z,t) + \Gamma \gamma y''(z,t) + d\chi.
    \end{equation}

The NNLEGI model is obtained from (\ref{eq:ContinuousLEGI}) when $w$ is approximated by a second-order finite difference approximation to (\ref{eq:y2DApproximation}), 
    \begin{equation}\label{eq:2d2ndorderfd}
       w^{(2,2)}(z) := \delta(z+a) - 2\delta(z) + \delta(z-a),
    \end{equation}
where $a = 1/m$ represents the size of each cell in the model, and $\delta$ the Dirac delta. This approximation is effectively dictating that each cell can only communicate with its two adjacent cells per unit time, however this modeling assumption need not hold, and we investigate in this paper the effect of assuming various ranges of cell communication per unit time. We will denote by {\it radius of influence} the number of cells involved in the communication process, per unit time. 
  Alternatively, we could use the kernel
    \begin{equation} \label{eq:2d4thorderfd} \scriptstyle
       w^{(2,4)}(z) := \frac{1}{12}\left(-\delta(z+2a) + 16\delta(z+a) - 30\delta(z) + 16\delta(z-a) - \delta(z-2a)\right),
    \end{equation}
which represents a fourth order approximation to the second derivative and effectively couples 5 cells per model time step, rather than 3. This is a higher-order approximation to the second derivative.
 Another way to increase the precision of our approximation is obtained by retaining higher order terms in the Taylor expansion \ref{eq:TaylorExpansion} and using the kernel 
    \begin{equation} 
    \label{eq:4d2ndOrderfd}\scriptstyle
        w^{(4,2)}(z) := -\delta(z+2a) + 4\delta(z+a) - 6\delta(z) + 4\delta(z-a) - \delta(z-2a),
    \end{equation}
which eliminates the second derivative term yet approximates the convolution kernel to fourth order. Each of these approximations highlight important immediate physical characteristics that must be taken into account during the modeling process: radius of influence and the characteristic of the derivative operator they are replacing. Our analysis will show that consideration of both the manner in which the convolution is approximated and the radius of influence play a role in the gradient sensing outcomes of the particular model. 

Invoking these approximations leads to a discretization of (\ref{eq:ContinuousLEGI}) similar to (\ref{eq:DiscreteOriginalLEGI}) except with the generalization,
\begin{equation*}
    dy_n = \beta r_n dt - \nu \sum_{n'=1}^m W^{(p,q)}_{nn'} y_{n'}dt + d\chi_n,
\end{equation*}
the matrix $W^{(p,q)}_{nn'}$, has parameters $p, q$ associated with the appropriate communication model $w^{(p,q)}$ 
described above (and appropriately-chosen boundary conditions). 

\section{Analysis of the General and Approximate Models} \label{sec:analysis}
We now turn our attention to understanding how the competition of deactivation, communication and thermal noise
plays out with an aim at better understanding the process of gradient sensing via intercellular communication. We focus on the radius of influence as well as the various alternative models for intercellular communication itself. 
\subsection{Competition between Deactivation and Communication}
 \label{sec:TransientDynamics}
We first examine the role played by deactivation and communication of the chemical $y$ on the dynamics transient dynamics. The key parameters are $\nu$ and $\gamma$. The analysis is easily performed, we only consider the mean dynamics, using Fourier transforms. The system (\ref{eq:ContinuousLEGI}) in Fourier space is
    \begin{equation*}
        \begin{aligned}
            \frac{\partial \hat{c}}{\partial t}(k, t) &= -k^2\hat{c} - \alpha \hat{c} + \mu \hat{r}, \\
            \frac{\partial \hat{r}}{\partial t}(k, t) &= \alpha \hat{c} - \mu \hat{r}, \\
            \frac{\partial \hat{x}}{\partial t}(k, t) &= \beta \hat{r} - \nu \hat{x}, \\
            \frac{\partial \hat{y}}{\partial t}(k, t) &= \beta \hat{r} - \nu \hat{y} - \gamma \hat{w}\hat{y},
        \end{aligned} 
    \end{equation*}
where $\hat{c}, \hat{r}, \hat{x}, \hat{y}, \hat{w}$ denote the Fourier counterparts to the original space variables. Note that $\hat{w}(k)$ is, by assumption, solely a function of the wavenumber $k$. This system of ordinary differential equations is linear, and we can represent it in matrix form as
    \begin{equation*}
        \frac{\partial}{\partial t}\hat{\mathbb{Y}}(k,t) = \mathcal{A}(k) \hat{\mathbb{Y}}(k,t), \qquad t>0.
    \end{equation*}
The solution of this equation is trivially given by
    \begin{equation}\label{eq:FourierSolution}
        \hat{\mathbb{Y}}(k,t) = e^{\mathcal{A}(k)t} \hat{\mathbb{Y}}(k, 0) = Pe^{\Lambda t}P^{-1}\hat{\mathbb{Y}}(k,0),  \quad t \ge 0,
    \end{equation}
where $\mathcal{A}(k) = P \Lambda P^{-1}$ is the diagonalization of $\mathcal{A}(k)$. In order to explore outcomes as a function of $\gamma$, $\nu$, and the choice of model for communication, we consider the evolution in time of the system subject to a Dirac delta distribution of concentration in $y$ at $t=0$. In this case, 
    \begin{equation} \label{eq:ySolutionFourier}
        y(z,t) = \frac{1}{\sqrt{2\pi}} \int_{-\infty}^\infty \hat{y}(k,0)e^{ikz}e^{-t(\gamma \hat{w}(k) + \nu)} dk.
    \end{equation}
From this expression, it is easy to see that the dispersion relation for this system of differential equations is given by 
    \begin{equation*}
        \omega(k) = i(\gamma \hat{w}(k)+\nu).
    \end{equation*}
We note that $\gamma \ge 0$ and $ \nu \ge 0$. Further the physically-relevant parameter regime is $\gamma > \nu$, which is to say that
the solution is highly dispersive, owing to the communication mechanism.
With the dispersion relationship in hand, it is easy to derive the phase and group velocities. These are, respectively,
    \begin{equation*}
        v_p = \omega(k)/k, \quad \mbox{ and} \quad  v_g = \frac{d\omega}{dk}.
    \end{equation*}
In what follows it is helpful to think of $ak$ as an estimate of the roughness of the external gradient. 
We note that $\hat w(0) = 0$, since mass is conserved. For small wave numbers, $\hat{w}(k)/k \rightarrow 0$ while $\nu/k\rightarrow \infty$, meaning that large wavelength disturbances are primarily associated with deactivation, associated with $\nu$, relative to the communication, associated with the parameter $\gamma$. For larger wave numbers, the communication term dominates the dynamics and do so in different ways, depending on the model. 
 The communication term improves the discernment of structure by the cell network, over not having a communication term. 
 
\begin{figure*}
        \centering
        \includegraphics[width=\textwidth]{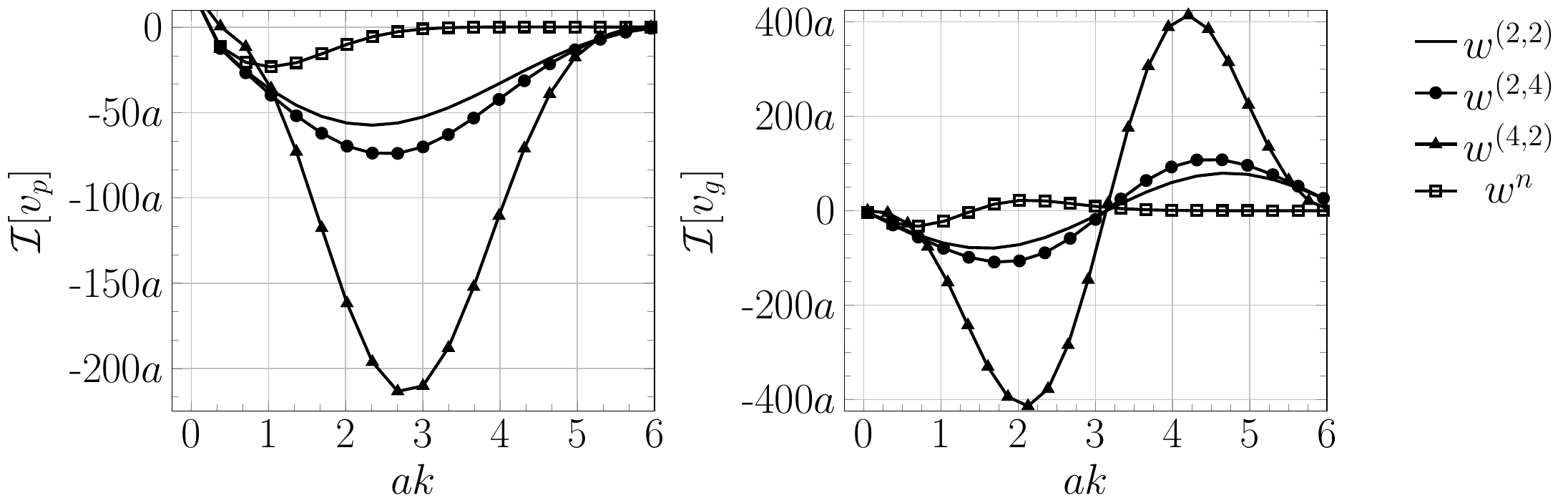}\caption{(a) Dispersion relation corresponding to three approximations, $w^{(2,2)}, w^{(2,4)}, w^{(4,2)}$, for the convolution term in (\ref{eq:ContinuousLEGI}), and for the Gaussian kernel $w^n$. The Nearest Neighbor model uses (\ref{eq:2d2ndorderfd}), (\ref{eq:2d4thorderfd}) uses a modified second derivative approximation, and finally (\ref{eq:4d2ndOrderfd}) includes the fourth derivative term to the approximation of (\ref{eq:ContinuousLEGI}). (b) Similarly, the group velocities for these approximations. Figures produced with parameters $\sqrt{\gamma/\nu} = 10$, smaller ratios decrease magnitude of velocities, especially for $ak>1$. }\label{fig:methodsdispersion}
\end{figure*}

For $\sqrt{\gamma /\nu}= 10 $, Figure \ref{fig:methodsdispersion} depicts the phase and group velocities the LEGI models associated with the various $w^{(p,q)}$. For completeness, 
we also consider the analysis of physical properties of a continuous kernel, derived from on a gaussian distribution,
\begin{equation}\label{eq:gaussiankernel}
 w^{n}(z) := \frac{d^2}{dz^2}\frac{ae^{-z^2/2a^2}}{\sqrt{2\pi}}.
\end{equation} 
We note that, in the large wavenumber regime, the phase velocity is monotonic for $w^n$, but not for the truncated approximations.

The phase and group velocity for the LEGI model solution, second derivative correction, fourth derivative correction along with the gaussian-inspired kernel are shown in Figure \ref{fig:methodsdispersion}(a) and Figure \ref{fig:methodsdispersion}(b), respectively.
We see from Figure \ref{fig:methodsdispersion} that the various models induce differences in qualitative and quantitative behavior. The qualitative analysis from these velocities will be critical to understanding the gradient precision calculations performed in section \ref{sec:SNR}.

We also see that both $w^{(2,2)}$ and $w^{(2,4)}$ are comparable quantitatively and qualitatively, and these in turn are different from the $w^{(4,2)}$ quantitatively. The figures suggest that the communications model is more critical than the radius of influence in modeling gradient sensing differences. 
There is a key regime of wavenumbers, wavenumbers corresponding to wavelengths that are smaller than the size of the cell and hence noisier, for which the phase velocities induced by the fourth derivative correction are much more negative than those induced by the second derivative approximations. The non-monotonicity of phase velocities create spatial interference for waves from that part of the spectrum, potentially decreasing the precision at which cells detect gradients. 

For the group velocity, the case is more nuanced: for very large wavelengths, the velocities induced by the fourth order correction are somewhat smaller than the other methods, while for smaller wavelengths this correction induces much larger velocity discrepancies. The difference between second derivative approximations are comparatively small for most wavenumbers. Again, we see the non-monotonicity of group velocities from the derivative approximations should cause interference among wavenumbers greater than the cell size, inducing noisy communication. The velocities of the gaussian-inspired kernel are not only smaller in magnitude, but also qualitatively distinct in that they decay much quicker than the other kernels. This decay should reduce the interference from the large wavenumber spectrum. Communication can make gradient discernment possible, particularly if the concentration gradient has a lot of structure at small scales, however, the manner in which this effect is modeled can have significant impact when comparisons are made between model outcomes and observations. From this analysis, we then expect that the higher-order derivative approximations will degrade the ability of a cell's sensory precision (which we will measure explicitly using the signal-to-noise ratio) and that the gaussian-inspired kernel will result in a much higher precision.

The dispersion relation suggests that the competition of the deactivation and communication  in this LEGI family of models 
leads to a non-trivial transient aspect to  gradient sensing and transport. The following calculation looks deeper on this competition in the transient phase. The analysis will focus on the second order model truncation, which can be carried farther analytically, but it is clear from what follows that the fourth order differential operator approximation would yield to a story with some similarities.
Assuming $\hat{w}(k)$ is sufficiently smooth, we can represent it in terms of a Taylor expansion about a cell location, which we truncate at the quadratic term. Keeping with our original assumptions about $w$ ($w$ is even) then we can simplify this series and write the stationary-phase solution as
    \begin{equation} \label{eq:ysolutionFourierApproximation}
        y(z,t) \approx \frac{e^{-t\nu}}{2\pi} \int e^{-i(kz - i\gamma k^2/2\hat{w}''(0)t)} dk,
    \end{equation}
for which a closed form solution for an approximation of $y(z,t)$ exists, {\it viz},
    \begin{equation} \label{eq:ysolutionApproximation}
        y(z,t)  \approx \frac{e^{-t\nu - \frac{z^2}{2\hat{w}''(0)\gamma t}}}{\sqrt{2\pi \hat{w}''(0) \gamma t}}, 
    \end{equation}
where $\hat w''(0)$ is understood to be the coefficient of the second order term in $\hat w$. Having this expression enables a fruitful analysis of the chemical's dependence on the parameters $\nu, \gamma$. 

There are three immediate conclusions that can be drawn from this formula. First, for a fixed $t$, $\gamma$ affects the variance of the distribution of $y$. This matches the intuition and physical reasoning that $\gamma$, being the communication parameter, diffuses the chemical $y$ among the cells in the system. Similarly, $\nu$ acts only in as a rate of exponential decay. This again intuitively matches are expectation that $\nu$ controls the rate of deactivation of $y$. Lastly, when $t$ is relatively small, the importance of the diffusive term, and thus, $\gamma$ and the choice of communication model, is greater than $\nu$. Setting the argument of the exponential in (\ref{eq:ysolutionApproximation}) to zero, it is possible to find a space/time horizon, which is approximately defined as $z/t = \pm \sqrt{\nu 2 \hat w''(0) \gamma}$. Taking this into account and the decaying nature of $y$ as a function of time it is thus expected that the autocorrelation of $y$ as a function of time starts compact, asymptotes to zero for large distances, and relaxes in time.

\begin{figure}
    \centering
    \includegraphics[width=0.48\textwidth]{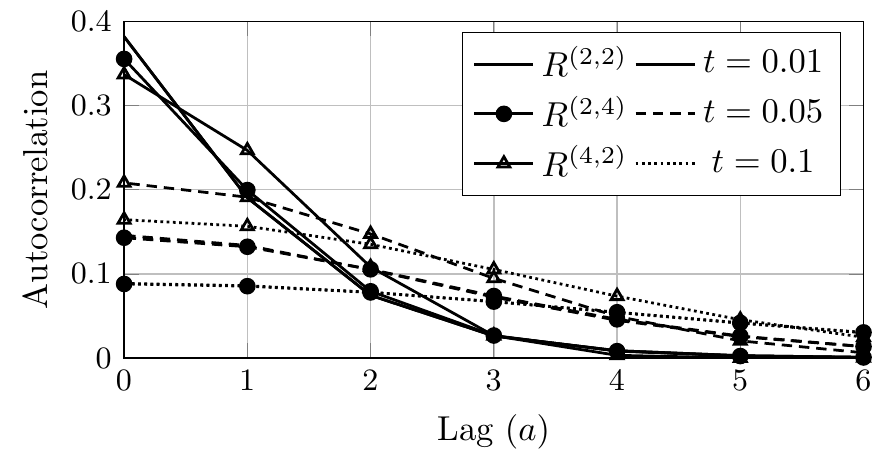}\caption{Spatial autocorrelation $R^{(p,q)}(z,t)$ of (\ref{eq:ySolutionFourier}) for the three communication kernels (\ref{eq:2d2ndorderfd}), (\ref{eq:2d4thorderfd}), (\ref{eq:4d2ndOrderfd}), respectively. For each $R^{(p,q)}(z,t)$ the snapshots of the decaying autocorrelations are taken at times $t = 0.01$ (highest $R$ at zero lag), $0.05$, and $0.1$ (lowest $R$ at zero lag). Autocorrelations of the two second derivative approximations are nearly indistinguishable for $t \geq 0.05$. Here $\sqrt{\gamma/\nu} = 10$.}
    \label{fig:autocorr}
\end{figure}

The spatio-temporal coupling that results in the competition of the deactivation and communication is quantified
in Figure \ref{fig:autocorr}, which shows the auto-correlation $R^{(p,q)}$ of $y$, for the various models. The expression (\ref{eq:ySolutionFourier}) was used in these calculations. It is seen that $R^{(2,2)}$ and $R^{(2,4)}$ are comparable in space and time, again, the radius of influence having little effect, when compared to the nature of the model: the $R^{(4,2)}$ has a much larger lag for a given time. We conclude that the fourth derivative correction induces greater ratio of communication to deactivation compared to both second derivative approximations. Furthermore, the fourth order model will convey a more dramatic signaling story if the external concentration gradient is changing quickly in time.

\subsection{Sensitivity and the Signal-to-Noise Ratio}\label{sec:SNR}
Having focused on the competition of deactivation and communication, we now consider how these are affected by the presence of noise intrinsic and extrinsic to each cell in the network. In \cite{Mugler2016} it is argued that gradient sensing for a given network could be assessed by computing the signal-to-noise ratio (SNR). The environment is noisy and the manner in which the deactivation effect modifies a signal is straightforward, hence, comparisons of SNR can be used to discern how effective communication is to gradient sensing. Presumably it is also a quantity that could lend itself to comparisons using laboratory data. Here we take this further by showing how the choice of model for communication affects the SNR.

To begin to understand the impact of the generalization of the communication term we give the mean steady state solutions to (\ref{eq:ContinuousLEGI}). Assuming appropriate boundary conditions, we denote $\overline{c}$ as the steady state solution for the concentration $c$, which satisfies $\nabla^2 c = 0$. A simple form for the concentration $c$ that obeys this equation and boundary conditions varies linearly in space with slope $g$ such that $ag/\overline{c} \ll 1$. We will assume this form for $\overline c$. The remaining steady-state solutions are:
    \begin{equation}
        \label{eq:SteadyState}
        \begin{aligned}
        \overline{r} &= \frac{\alpha}{\mu} \overline{c}, \\
        \overline{x} &= \frac{\beta}{\nu} \overline{r} = \frac{\beta \alpha}{\mu \nu} \overline{c}, \\
        \overline{y} &= \mathcal{F}^{-1} \biggr\lbrace \frac{\frac{\beta}{\nu} \mathcal{F}(\overline{r})}{1+ \frac{\gamma}{\nu} \mathcal{F}(w)} \biggr\rbrace \\
        &= \frac{\alpha \beta}{\mu \nu} \mathcal{F}^{-1} \biggr\lbrace \frac{ \mathcal{F}(\overline{c})}{1+ \frac{\gamma}{\nu} \mathcal{F}(w)} \biggr\rbrace,
        \end{aligned}
    \end{equation}
where $\mathcal{F}$ is the spatial Fourier transform defined as
    \begin{equation}\label{eq:FourierTransform}
    \mathcal{F}(g) = \frac{1}{\sqrt{2\pi}} \int_{-\infty}^{\infty} g(z) e^{-ikz}dz.
    \end{equation}
Now, turning to the discrete setting for the rest of this section, the last line in (\ref{eq:SteadyState}) becomes
\begin{equation}
    \label{eq:DiscreteSteadyState}
    \overline{y}^r_n = \frac{\alpha \beta}{\mu\nu} \sum_{n'=1}^{m} K_{nn'}^{(p,q)}\overline{c}_{n'}.
\end{equation}
where $K_{nn'}^{p,q} = \left(W^{p,q}\right)^{-1}_{nn'}$. When $\gamma=0$, there is no communication, and $\overline{y} = \overline{x}$. When the communication term is non-zero, the deviation of $\overline{y}$ from $\overline{x}$ provides an indication of the impact of communication on the system and if the cell is near higher concentrations of $c$. To this end, we follow \cite{Mugler2016} and define the following deviation of mean states,
    \begin{equation}\label{eq:Deltabar}
        \overline{\Delta}_n = \left(\overline{x}_n - \overline{y}_n\right).
    \end{equation}
The ability for a cell to detect a chemical gradient among a noisy chemical background is the ratio of the square of this mean deviation and the variance in chemical concentrations of $x, y$, 
    \begin{equation}\label{eq:SNR}
        \textrm{SNR} = \left(\frac{\overline{\Delta}_n}{\delta \Delta_n}\right)^2,
    \end{equation}
where $(\delta \Delta_n)^2 = \left(\delta x_n\right)^2 + \left(\delta y_n \right)^2 - 2 C^{xy}_{nn'}$, and $C^{xy}_{nn'}$ is the cross-correlation between $x$ and $y$. To calculate the variances, we start by computing the power spectra of $r, x$ and $y$: $S^{rr}, S^{xx}$, and $S^{yy}$. We must make a series of assumptions for the relevant time scales: the integration time $T$ is longer than the receptor equilibration time, messenger turnover time ($1/\nu$), and messenger exchange time ($1/\gamma$) (we borrow these assumptions from \cite{Mugler2016}, which describe and reason them in more detail). Under these assumptions, covariances in long-time averages are given by the low frequency limits of the power spectra, $C^{xy}_{nn'} = \lim_{\omega\rightarrow 0} S^{xy}_{nn'}(\omega)$. In the discrete setting \cite{Mugler2016} invokes the fluctuation-dissipation theorem, makes a linear noise approximation, and provides a derivation of $S^{rr}_n$, which we use unaltered in our generalization,
\begin{equation}
    \label{eq:EtaPowerSpectrum}
    S^{rr}_{nn'}(\omega) = \frac{2\alpha \overline{c}_n}{\mu^2} \begin{cases}
    (1+\frac{\alpha}{2\pi a D} & n' = n\\
    \frac{\alpha}{4\pi a D} \frac{1}{|n-n'|} & \neq n'\neq n.
    \end{cases}
\end{equation}
The power spectra $S^{xx}_{nn'}$, $S^{yy}_{nn'}$, and $S^{xy}_{nn'}$ can more easily be computed by calculating the (co)variances of the Fourier transforms of (\ref{eq:ContinuousLEGI}) in space and time,
    \begin{equation} \label{eq:PowerSpectra}
        \begin{aligned} \scriptstyle
        S_{nn'}^{xx}(\omega) &= 
           \scriptstyle\frac{1}{\nu^2 + \omega^2}\left(S_{nn'}^{rr}(\omega) + \langle \mathcal{F}(d\xi_n)^* \mathcal{F}(d\xi_{n'})\rangle \right) , \\
         \scriptstyle S_{nn'}^{yy}(\omega) &= 
         \scriptstyle \frac{1}{\nu^2}\sum_{jj'}\tilde{K}_{nj}^{p*}\tilde{K}_{n'j'}^p \left(S_{jj'}^{rr}(\omega) +  \scriptstyle\langle \mathcal{F}(d\chi_j)^* \mathcal{F}(d\chi_{j'})\rangle \right),\\
         \scriptstyle S_{nn'}^{xy}(\omega) &=  
         \scriptstyle\frac{1}{\nu(\nu+i\omega)}\sum_{j}\tilde{K}_{nj}^pS^{rr}_{n'j}(\omega)
        \end{aligned}
    \end{equation}
where $\tilde{W}_{nn'} := W_{nn'} - i (\nu/\omega)\delta_{nn'}$. The spectra of thermal noise terms $d\xi_n$ and $d\chi_n$ can be calculated using the Fourier transform in a similar way and using the steady-state means to get, respectively,
    \begin{equation} \label{eq:NoisePowerSpectra}
    \begin{aligned}
    \langle \mathcal{F}(d\xi_n)^* \mathcal{F}(d\xi_{n'})\rangle &= 2\nu \overline{x}_n, \\
    \langle \mathcal{F}(d\chi_n)^* \mathcal{F}(d\chi_{n'})\rangle &= 2\nu W^{(p,q)}_{nn'}\overline{y}_{n'}. \\
    \end{aligned}
    \end{equation}

\begin{figure}
    \begin{center}
    \includegraphics[width=0.49\textwidth]{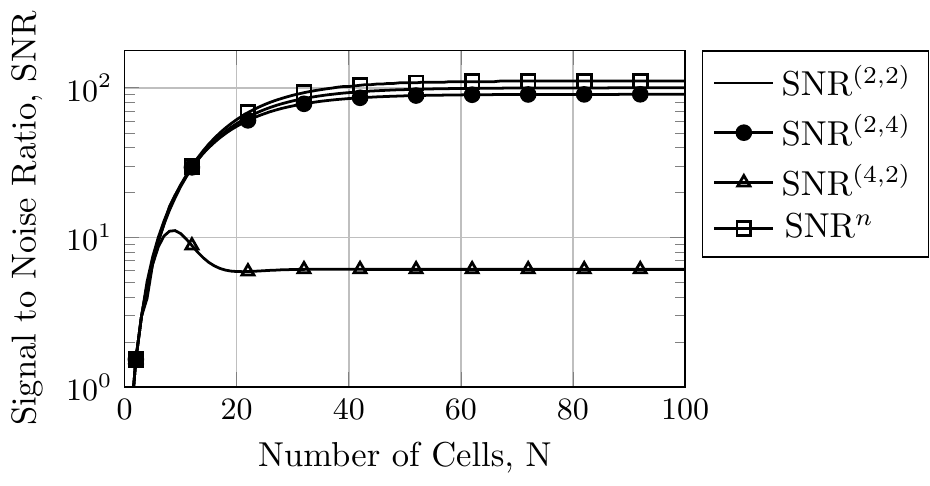}\caption{Precision of gradient sensing with temporal integration. Signal to noise ratio SNR$^{(p,q)}$, defined as (\ref{eq:SNR}), when using approximations (\ref{eq:2d4thorderfd}) SNR$^{(2,4)}$ and (\ref{eq:4d2ndOrderfd}) SNR$^{(4,2)}$ as compared to the original communication model (\ref{eq:2d2ndorderfd}) SNR$^{(2,2)}$. Here, $N$ refers to the number of cells in the system. We use $\alpha/a^3 \mu = \beta/\nu = 100$, $D = 50\mu$m$^2$/s, $a = 10\mu$m, $\mu = \nu = 1$s$^{-1}$, $T = 10$s, and $\sqrt{\gamma/\nu} = 10$. Parameters values as suggested in \cite{Mugler2016}}\label{fig:SNR}
    \end{center}
\end{figure}

Finally, combining (\ref{eq:PowerSpectra}) into the definition of $(\delta \Delta_n)^2 = (\delta x_n)^2 + (\delta y_n)^2 - 2C^{xy}_{nn}$ and (\ref{eq:SNR}) allows us to compute a metric for the ability of the cells to extract gradient information from a noisy chemical background. In Figure \ref{fig:SNR}, we depict the SNR corresponding to the various approximations to $w^{(p,q)}$: (\ref{eq:2d2ndorderfd}), (\ref{eq:2d4thorderfd}), and (\ref{eq:4d2ndOrderfd}). These yield the curves $\textrm{SNR}^{(2,2)}$,
$\textrm{SNR}^{(2,4)}$, and $\textrm{SNR}^{(4,2)}$, respectively, as a function of system size.

For large cell numbers, the precision of gradient sensing when including more accurate approximations to the convolution seemingly degrages. We also see that the Gaussian-inspired kernel induces the greatest sensory precision of the models tested. The radius of influence is not the critical aspect, since each of these kernels have radii larger than the nearest neighbor kernel; rather, the analytic properties associated with each parameterization are more aligned with model outcomes. The (\ref{eq:4d2ndOrderfd}) model does not share all of the same qualitative characteristics with the other models. We find that the use of the fourth derivative approximation (\ref{eq:4d2ndOrderfd}) has a dramatic impact on the ability of cells to determine a chemical gradient. For large system size, and a high order approximation of the kernel, the gradient sensing degrades by an order of magnitude. The impact of the more effective communication model (see Figure \ref{fig:autocorr}) is to propagate more information from noisy frequencies. 

\section{Summary and Conclusions}
\label{sec:disc}
Gradient sensing in chains of cells and organisms that have no specialized gradient sensing organelles is not well understood. Experimental evidence lends support to the conjecture that the ability of chains of cells to sense the gradient of an external chemical concentration could rest upon cell-to-cell communication. This hypotheses is central to a Local Excitation Global Inhibition (LEGI) model proposed in Mugler \textit{et al.}, \cite{Mugler2016}. In this particular LEGI model, communication is restricted to nearest neighbor communication and thus we denote it as the Nearest Neighbor LEGI (NNLEGI) model. Mugler \textit{et al.} succeeded in demonstrating that gradient sensing was possible with the NNLEGI model. In this paper we replaced the nearest neighbor communication term in the NNLEGI model by a convolution and thus generalized their model. By generalizing we aimed to explore how the gradient sensing capabilities of the model depends on the radius of influence as well as on the approximating order of the convolution kernel itself. The NNLEGI model is shown to be subsumed into the more general model.

The radius of influence refers to the number of neighbors involved in the communication per unit time. The approximating order refers in turn on how the convolution kernel is approximated locally. Understanding how these impact gradient sensing, it is argued here, could lead to better models for gradient sensing, but more importantly, to a better basic understanding on how cells and collections of these sense an external gradient. 
 
When approximating communication via convolution in a dynamics model representing LEGI it was found that the manner in which information was shared was more critical than on the radius of influence. The generalized NNLEGI model was analyzed with the aim of establishing how cell size, cell chain length, and alternative proposals for the inter-cellular communication mechanisms manifest themselves in the gradient sensing model outcomes. We found that the choice of communication parameterization produced meaningful differences on model outcomes like information velocities (Figure \ref{fig:methodsdispersion}), signal auto-correlation (Figure \ref{fig:autocorr}), and the SNR (Figure \ref{fig:SNR}). Further, we determined spatio-temporal ranges where one expects to see deactivation-dominated or advective-dominated (via communication) reactions and how the sensitivity of the model was related to those parameters and choice of communication kernel.

In \cite{Mugler2016} the signal-to-noise ratio (SNR) was used as a way to assess the relative importance of the local excitation to the global inhibition effects on gradient sensing, the latter effect captured via communication. We found that the NNLEGI exhibits high SNR as compared to a model equipped with higher order approximations for the 	communication. Moreover, the NNLEGI and the continuous Gaussian-inspired kernel mode for communication were
qualitatively comparable. The dispersion relation of the second order and the fourth order models were similar and these were not similar to the continuous Gaussian-inspired kernel.
If the environment is quiet gradient sensing was more effective in the higher order model approximation of communication. The higher order approximate model, however, exhibits a significantly lower SNR as a consequence of more effective communication and propagation of higher frequency information. As a result, it is more prone to propagate thermal noise.

It can be argued that the generalization of communication via convolution is a general way to explore how communication can play a role in gradient sensing networks. Its generalization lends itself to a higher level of abstraction we think is useful in the consideration of networks in higher dimensions as well as networks with inherent heterogeneity. The following analysis might prove useful in constructing models that capture and explain laboratory 
observations. The analytical tools introduced in this paper can play an even more critical role in understanding communication pathways in these more complex networks and how these pathways are affected by local reactions and noise. 

\section*{Acknowledgements}\label{sec:Acknowledgements}
We thank A. Mugler and B. Sun for fruitful discussions.
JMR wishes to thank the Kavli Institute of Theoretical Physics (KITP), at the University of California, Santa Barbara for their hospitality and for supporting his research on this project. The KITP is supported in part by the National Science Foundation under Grant No. NSF PHY-1748958. CV and BF were Research Experiences for Undergraduate participants at Oregon State University in summer of 2017. The REU program is sponsored by the National Science Foundation under grant No. NSF DMS-1359173.

\bibliography{chemotax_APS}

\end{document}